\documentclass[twocolumn,twoside]{IEEEtran}

\ifCLASSINFOpdf

\else

\fi

\ifCLASSOPTIONcompsoc
\usepackage[caption=false, font=normalsize, labelfont=sf, textfont=sf]{subfig}
\else
\usepackage[caption=false, font=normalsize]{subfig}
\fi
\usepackage{lipsum}%
\usepackage[dvipsnames]{xcolor}

\usepackage{multicol}   
\usepackage{cite}
\usepackage{gensymb}
\usepackage{multirow}
\usepackage{graphics}
\usepackage{epsfig}
\usepackage{graphicx}
\usepackage{epstopdf}
\usepackage{textcomp}
\usepackage{amsmath}
\usepackage{mathtools}
\usepackage{filecontents}
\usepackage{lipsum,color}
\usepackage{amssymb}
\usepackage{float}
\usepackage{comment}

\usepackage{amsthm}

\usepackage{braket}
\usepackage{times} 
\usepackage{amsthm}  

\usepackage{amsfonts}

\theoremstyle{break}
\begin{document}
\raggedbottom
\makeatletter
\def\fps@figure{!t}   
\def\fps@table{!t}    
\makeatother

\title{CV Quantum Communications with Angular Rejection Filtering: Modeling and Security Analysis}

\author{Mohammad~Taghi~Dabiri,~Meysam~Ghanbari,
	~Rula~Ammuri,
	~Saif~Al-Kuwari,~{\it Senior Member,~IEEE}, \\
	~Mazen~Hasna,~{\it Senior Member,~IEEE}, 
	~and~Khalid~Qaraqe,~{\it Senior Member,~IEEE}

	\thanks{M.T. Dabiri, M. Ghanbari, and S. Al-Kuwari are with the Qatar Center for Quantum Computing, College of Science and Engineering, Hamad Bin Khalifa University, Doha, Qatar. email: (mdabiri@hbku.edu.qa; megh89467@hbku.edu.qa; smalkuwari@hbku.edu.qa).}
	
	\thanks{Rula Ammuri is with Professionals for Smart Technology (PST), Amman, Jordan (email: rammuri@pst.jo).}
	
	\thanks{Mazen Hasna is with the Department of Electrical Engineering, Qatar University, Doha, Qatar (e-mail: hasna@qu.edu.qa).}
	
	\thanks{Khalid A. Qaraqe is a professor with the College of Science and Engineering, Hamad Bin Khalifa University, Doha, Qatar, and  an adjunct  professor with the Department of Electrical Engineering, Texas A\&M University at Qatar, Doha, Qatar (e-mail: kqaraqe@hbku.edu.qa)}
	
	\thanks{This publication was made possible by NPRP14C-0909-210008 from the Qatar Research, Development and Innovation (QRDI) Fund (a member of  Qatar Foundation). } 
}

\maketitle
\begin{abstract}
Continuous-variable quantum key distribution (CV-QKD) over free-space optical links is a promising approach for secure communication, but its performance is limited by turbulence, pointing errors, and angular leakage that can be exploited by an eavesdropper. To mitigate this, we consider an angular rejection filter that defines a safe-zone at the receiver and blocks signals from outside the desired cone. A system and channel model is developed including turbulence, misalignment, and safe-zone effects, and information-theoretic metrics are derived to evaluate security. Simulation results show that the safe-zone significantly reduces information leakage and that careful tuning of beam waist, angular threshold, and aperture size is essential for maximizing the secret key rate. Larger apertures improve performance but increase receiver size, while longer links require sub-100~$\mu$rad alignment accuracy. These results highlight safe-zone enforcement and parameter optimization as effective strategies for practical and secure CV-QKD.
\end{abstract}

%

%
\IEEEpeerreviewmaketitle


\section{Introduction}
Two principal paradigms realize quantum key distribution (QKD): discrete-variable (DV) and continuous-variable (CV). DV encodes bits in discrete photonic states (polarization/time-bin) and requires single-photon detectors, whereas CV-QKD uses Gaussian-modulated quadratures of coherent states measured via homodyne/heterodyne with a local oscillator \cite{Cao2022}. This shift from photon counting to coherent reception enables reuse of telecom hardware, supports higher symbol rates, enables room-temperature operation, and facilitates integration with coherent/WDM optical networks \cite{Moreolo2025}. These advantages make CV-QKD attractive for free-space optical (FSO) deployments, including ground-to-ground urban links, UAV relays, and satellite communications, where short contact windows, strict SWaP constraints, and strong background illumination are critical \cite{Hosseinidehaj2017}.

Despite these advantages, CV-QKD performance in FSO links is strongly affected by channel impairments, primarily atmospheric turbulence and pointing errors. Turbulence arises from refractive index fluctuations in the atmosphere, producing beam wander, scintillation, and wavefront distortions that introduce excess noise and degrade the secret key rate (SKR) \cite{Li2020}. Pointing errors originate from platform vibrations, beam jitter, or tracking inaccuracies, which reduce collected power and increase channel loss, leading to substantial SKR degradation \cite{Alshaer2021}.  

A large body of work has modeled and mitigated these impairments. For turbulence, \cite{Ghalaii2022} moved beyond weak-fluctuation models using log-normal and extended Huygens–Fresnel descriptions, revealing severe SKR limits for near-horizon satellite links. Later, \cite{Ghalaii2023} analyzed CV-MDI-QKD under turbulence (Rytov variance; Fried’s coherence length) and showed that centroid fluctuations and spot-size growth strongly reduce transmissivity even in weak–moderate regimes. In MIMO settings, \cite{Kumar2025} modeled turbulence-induced fading with log-normal statistics, exposing scalability constraints, though aperture diversity can help. Beyond air, underwater CV-QKD has been studied in \cite{Meena2025}, where turbulence from salinity/temperature gradients plus absorption and scattering severely limits range unless advanced techniques (e.g., virtual photon subtraction) are used.  

To counter turbulence, several methods have been proposed. A rate-adaptive reconciliation protocol \cite{Gumus2025} adapts coding rates to instantaneous SNR, boosting reconciliation efficiency by more than 150\%. In \cite{Miao2025}, generalized Kennedy receivers with dynamic displacement improved robustness compared to homodyne detection in log-normal fading. Phase-sensitive amplifiers (PSAs) placed before homodyne detection were shown in \cite{Alshaer2024} to compensate turbulence-induced noise and extend secure distances. These studies confirm the feasibility of enhancing turbulence resilience using coding, receiver design, and optical amplification.  

Pointing error has also been modeled extensively. In \cite{Dequal2021}, a statistical model showed that microradian-level beam displacement induces fading and excess noise that drastically lower SKR. In multiuser satellite settings, \cite{Phan2023} derived closed-form relations between pointing error loss, QBER, sifted key probability, and SKR. Recent works also explored mitigation: experimental acquisition, pointing, and tracking (APT) systems with fast steering mirrors demonstrated stable CV-QKD under daylight turbulence \cite{Zheng2025}. Adaptive beam shaping was proposed in \cite{Dabiri2025} to optimize divergence and intensity profiles, reducing jitter-induced losses. Receiver-side aperture optimization in \cite{Nguyen2023} balanced pointing tolerance with background noise suppression, while UAV-based CV-QKD with dual-polarization QPSK was studied in \cite{Alshaer2022}, showing that optimized divergence, field of view (FoV), and transmit power can tolerate centimeter-scale boresight displacements while sustaining SKR.  

Motivated by these gaps, this paper develops an integrated framework for CV-QKD in FSO channels with safe-zone enforcement. We present a tractable system and channel model that jointly captures turbulence, pointing errors, and safe-zone effects; derive information-theoretic expressions for mutual information, the Holevo bound, and SKR under these impairments; introduce an angular rejection filter–based safe-zone model that explicitly quantifies information leakage to Eve; and, through analysis and simulations, demonstrate the critical role of optimally tuning the beam waist, angular threshold, and aperture size. The results show that larger apertures and optimized thresholds enhance security but incur size/complexity trade-offs, while longer links demand sub-100~$\mu$rad alignment accuracy to sustain positive SKR. Taken together, these findings highlight safe-zone enforcement and parameter optimization as practical strategies for secure and scalable CV-QKD deployment in free-space networks.

\begin{figure}
	\begin{center}
		\includegraphics[width=2.7 in]{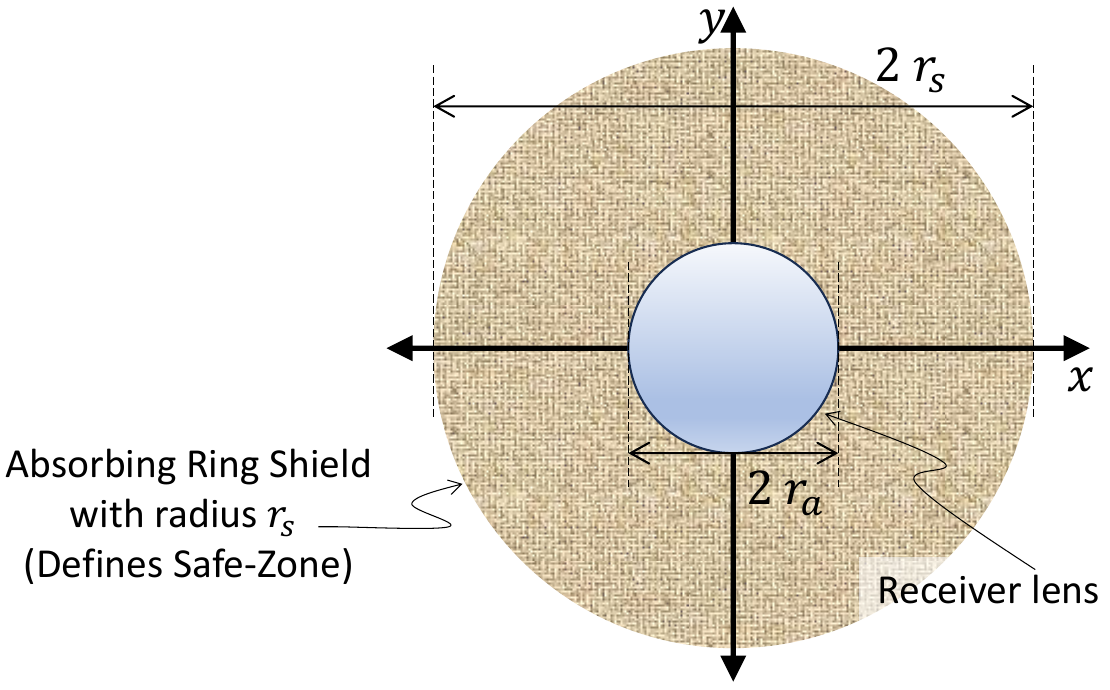}
		\caption{
			Illustration of the receiver configuration with an \emph{Angular Rejection Filter} placed behind the main lens. This absorbing ring-shaped shield defines the safe-zone by blocking and absorbing optical signals arriving from outside the desired angular cone (with half-angle $\theta_{\text{safe}}$), thereby preventing them from reaching the quantum detector.
		}
		
		\label{sm1}
	\end{center}
\end{figure}

\section{System Model}

\subsection{System Overview}
We consider a continuous-variable quantum communication link between a ground-based transmitter (Alice) and a receiver, located either on a UAV or on the ground, hereafter referred to as Bob. The quantum information is encoded using Gaussian-modulated coherent states transmitted through a free-space optical (FSO) channel. The receiver system is equipped with a circular aperture lens and an \emph{Angular Rejection Filter} (ARF) positioned behind the lens to physically enforce an angular \emph{safe-zone} defined by a half-angle $\theta_{\text{safe}}$ as shown in Fig. \ref{sm1}. 

The optical link is impaired by two main physical effects: atmospheric turbulence and random pointing errors due to UAV motion and tracking inaccuracies. Both effects cause stochastic variations in the received signal power. These are modeled by multiplicative channel transmissivities denoted by $\eta_B$ for Bob and $\eta_E$ for Eve, each comprising a turbulence and pointing component. For security evaluation, we adopt a worst-case assumption in which the eavesdropper is ideal and experiences no atmospheric turbulence or fading. Specifically, we assume that Eve is located outside the safe-zone and can collect the entire fraction of optical power that geometrically leaks beyond the angular rejection filter, as determined by the pointing deviation of the Gaussian beam. Thus, the transmissivity to Eve, $\eta_E$, is determined solely by the angular misalignment relative to the safe-zone boundary and does not include any turbulence-related attenuation, i.e., $\eta_{\text{tur}}^{(E)} = 1$.

\subsection{Continuous-Variable Quantum Communication}
In CV-QKD, the quantum information is encoded onto the quadratures of coherent states via Gaussian modulation. A coherent state $\left| \alpha \right\rangle$ is defined by a complex amplitude $\alpha = q + i p$, where $q$ and $p$ are real-valued quadratures corresponding to the position and momentum operators. In our system, Alice generates these coherent states by modulating a laser beam such that \cite{Alshaer2024}:
\begin{align}
	q, p \sim \mathcal{N}(0, V_m)
	\label{eq:modulation}
\end{align}
where $V_m$ is the modulation variance expressed in shot-noise units (SNU).

The resulting physical signal corresponds to a coherent quantum state whose quadrature operators include both the modulated classical values and the intrinsic vacuum noise. Specifically, the input quadratures can be expressed as:
\begin{align}
	\hat{q}_{\text{in}} = q + \hat{q}_v, ~~\&~~
	\hat{p}_{\text{in}} = p + \hat{p}_v
	\label{eq:input_quadratures}
\end{align}
where $\hat{q}_v$ and $\hat{p}_v$ are independent vacuum noise operators with zero mean and unit variance. As a result, the total quadrature distributions of the input state remain Gaussian with variance $V_m + 1$.

\subsection{Transmitted Optical Signal Model}
The quantum signal transmitted by Alice is a modulated coherent state obtained by applying Gaussian-distributed quadrature displacements to a continuous-wave laser beam. Let the carrier frequency be denoted by $f_0$, corresponding to an optical wavelength of approximately 1550~nm. 
For conceptual clarity, we represent the modulated optical field in the time domain as \cite{Alshaer2024}:
\begin{align}
	E_{\text{sig}}(t) \propto q \cos(2\pi f_0 t) + p \sin(2\pi f_0 t)
	\label{eq:optical_field}
\end{align}
where $(q, p)$ are independent Gaussian random variables with variance $V_m$ in shot-noise units (SNU). This normalized representation abstracts away absolute scaling factors (e.g., photon energy or optical power) and emphasizes that the transmitted field is a stochastic process with Gaussian statistics, zero mean, and total variance $V_m + 1$ per quadrature in SNU.

\subsection{Received Signal Model at Bob and Eve}
The quantum signal transmitted by Alice is subject to attenuation and noise as it propagates through the wireless optical channel. The impairments include free-space path loss, atmospheric turbulence, and random pointing errors. These effects are captured through two effective transmissivity coefficients: $\eta_B$ for the legitimate receiver (Bob) and $\eta_E$ for the potential eavesdropper (Eve). Each of these can be expressed as the product of two independent components:
\begin{align}
	\eta_B = \eta_{\text{po}}^{(B)} \cdot \eta_{\text{tur}}^{(B)}, \quad
	\eta_E = \eta_{\text{po}}^{(E)} \cdot \eta_{\text{tur}}^{(E)}
	\label{eq:transmissivities}
\end{align}
At Bob's receiver, the modulated quantum field is converted to an electrical signal using coherent detection. Assuming homodyne detection of the $q$ quadrature for simplicity, the measured signal at Bob is modeled as:
\begin{align}
	y_B = \sqrt{\eta_B} \cdot \hat{q}_{\text{in}} + \hat{n}_B
	\label{eq:received_bob}
\end{align}
where $\hat{n}_B$ denotes the equivalent additive noise at Bob, including vacuum noise from loss and detector noise. The total noise variance observed at Bob is given by:
\begin{align}
	\sigma_B^2 = (1 - \eta_B) + \xi
	\label{eq:noise_bob}
\end{align}
where $\xi$ represents the excess noise (e.g., due to background radiation or hardware imperfections) referred to the channel input.

Similarly, Eve's received signal (assuming she employs an ideal receiver) can be written as:
\begin{align}
	y_E = \sqrt{\eta_E} \cdot \hat{q}_{\text{in}} + \hat{n}_E
	\label{eq:received_eve}
\end{align}
where $\hat{n}_E$ includes the vacuum noise from Eve's loss channel. Although Eve's exact detection strategy may vary, this linear model provides a worst-case upper bound for security analysis.

\subsection{Information-Theoretic Metrics for Security Evaluation}
To evaluate the security performance of the system, we consider the standard information-theoretic quantities used in CV-QKD protocols. The first quantity is the mutual information between Alice and Bob, assuming a Gaussian-modulated coherent-state protocol and homodyne detection \cite{Derkach2020}:
\begin{align}
	I_{AB} = \frac{1}{2} \log_2 \left( 1 + \frac{\eta_B V_m}{\sigma_B^2} \right)
	\label{eq:mutual_info_ab}
\end{align}
where $\sigma_B^2$ is the total noise variance at Bob referred to the channel input modeled in \eqref{eq:noise_bob}, and $V_m$ is the modulation variance.

The second quantity is the Holevo bound $\chi_{AE}$, which quantifies the maximum information that Eve can extract from the quantum states sent by Alice. Under the assumption of a collective Gaussian attack and optimal detection by Eve, the Holevo information can be approximated as:
\begin{align}
	\chi_{AE} = S(E) - S(E|x)
	\label{eq:holevo}
\end{align}
where $S(E)$ is the von Neumann entropy of Eve's state, and $S(E|x)$ is the entropy conditioned on Alice’s modulation. For a pure-loss channel with transmissivity $\eta_E$, the Holevo quantity simplifies to \cite{Derkach2020}:
\begin{align}
	\chi_{AE} = \frac{1}{2} \log_2 \left( \frac{V_m + 1}{1 + (1 - \eta_E) V_m / (1 + \eta_E)} \right)
	\label{eq:holevo_simple}
\end{align}

Finally, the asymptotic secret key rate under reverse reconciliation is given by \cite{Derkach2020}:
\begin{align}
	K = \beta I_{AB} - \chi_{AE}
	\label{eq:skr}
\end{align}
where $\beta \in (0,1]$ is the reconciliation efficiency. These metrics allow us to evaluate how the pointing error, turbulence, safe-zone (captured through $\eta_B$ and $\eta_E$) affect the security of the quantum communication system.

\section{Channel Modeling}
In this section, we model the overall transmissivity of the wireless optical channel for both Bob and the Eve. The total channel transmissivity for Bob, denoted as \( \eta_B \), is modeled as the product of two independent components:
\begin{align}
	\eta_B = \eta_{\text{sys}} \cdot \eta_{\text{po}}^{(B)} \cdot \eta_{\text{tur}}^{(B)}
\end{align}
where \( \eta_{\text{po}}^{(B)} \) accounts for the stochastic attenuation due to beam misalignment caused by pointing errors, \( \eta_{\text{tur}}^{(B)} \) models the fading induced by atmospheric turbulence, and \( \eta_{\text{sys}} \in (0,1] \) is a deterministic system loss factor that captures constant optical losses due to hardware imperfections, alignment bias, and free-space path attenuation under nominal conditions.

\subsection{Atmospheric Turbulence}
The atmospheric turbulence is modeled using the Gamma-Gamma distribution, which is widely accepted for FSO channels in moderate to strong turbulence regimes. The probability density function (PDF) of the channel gain \( h \sim \text{GG}(\alpha,\beta) \) is given by \cite{dabiri2019tractable}:
\begin{align}
	f_h(h) = \frac{2(\alpha\beta)^{\frac{\alpha+\beta}{2}}}{\Gamma(\alpha)\Gamma(\beta)} h^{\frac{\alpha+\beta}{2}-1} K_{\alpha-\beta}\left(2\sqrt{\alpha\beta h}\right), \quad h > 0
\end{align}
where \( \alpha \) and \( \beta \) are the shaping parameters related to atmospheric conditions, \( \Gamma(\cdot) \) is the Gamma function, and \( K_{\nu}(\cdot) \) is the modified Bessel function of the second kind. The transmissivity due to turbulence is $\eta_{\text{tur}}^{(B)} = h$.

\subsection{Pointing Error Model}
We define a 3D coordinate system $(x,y,z)$ where the \( z \)-axis corresponds to the direct line-of-sight (LoS) path from the transmitter (Alice) to the center of Bob's receiving lens aperture. The receiver aperture is located in the plane \( z = Z_L \), and its center is located at the origin \((x=0, y=0)\) in the transverse plane.

Due to random tracking errors, the transmitted beam deviates from the central axis. The angular deviation is modeled as a two-dimensional Gaussian random variable \cite{dabiri2019tractable}:
\begin{align}
	\theta_{e_x}, \theta_{e_y} \sim \mathcal{N}(0, \sigma_{\theta}^2)
\end{align}
This angular error translates to a lateral displacement of the beam center in the \(x\)--\(y\) plane at distance \( Z_L \) given by:
\begin{align}
	x_e = Z_L \cdot \theta_{e_x}, \quad y_e = Z_L \cdot \theta_{e_y}
\end{align}

The Gaussian laser beam at the receiver plane has an intensity profile given by \cite{dabiri2019tractable}:
\begin{align}
	I(x, y) = I_0 \cdot \exp\left( -\frac{2 \left((x - x_e)^2 + (y - y_e)^2\right)}{w^2(Z_L)} \right)
\end{align}
where \( I_0 \) is the peak intensity and \( w(Z_L) \) is the beam radius at the distance \( Z_L \) \cite{Ghalaii2023}:
\begin{align}
	w(Z_L) = w_0 \sqrt{1 + \left( \frac{\lambda Z_L}{\pi w_0^2} \right)^2 }
\end{align}
Here, \( w_0 \) is the beam waist and \( \lambda \) is the laser wavelength.

The fraction of power collected by Bob's circular aperture of radius \( r_a \), considering the misalignment, is given by:
\begin{align}
	\eta_{\text{po}}^{(B)} = \int_{\text{Aperture}} \frac{I(x,y)}{\int_{\mathbb{R}^2} I(x,y) dx dy} dx dy
\end{align}
This results in the well-known expression for Gaussian misalignment fading \cite{dabiri2019tractable}:
\begin{align} \label{sbv1}
	\eta_{\text{po}}^{(B)} = A_0 \cdot \exp\left( -\frac{2 r_e^2}{w^2(Z_L)} \right)
\end{align}
where:
$r_e^2 = x_e^2 + y_e^2 = Z_L^2 (\theta_{e_x}^2 + \theta_{e_y}^2)$,  and $A_0 = \left( \text{erf}\left(\frac{\sqrt{2} r_a}{w(Z_L)}\right) \right)^2$.

\subsection{Power Partitioning and Safe-Zone Analysis}
To evaluate the security impact of spatial confinement, we analyze how the total optical power at the receiver plane is partitioned between the legitimate receiver and the eavesdropper. As shown in Fig. \ref{sm1}, the physically enforced safe-zone is implemented by combining the main receiving lens of radius $r_a$ with a surrounding circular absorbing guard, forming a total effective radius $r_{\text{safe}} = Z_L \cdot \theta_{\text{safe}}$ in the transverse plane. This defines a power-capturing region within which all incident light is considered secure.

Following \eqref{sbv1}, the total collected power within the safe-zone, including both the main lens and the surrounding guard ring, is given by:
\begin{align}
	\eta_{\text{safe}} = A_{\text{safe}} \cdot \exp\left( -\frac{2 r_e^2}{w^2(Z_L)} \right)
\end{align}
where $r_e^2 = Z_L^2 (\theta_{e_x}^2 + \theta_{e_y}^2)$ is the instantaneous squared radial pointing offset, and:
\begin{align}
	A_{\text{safe}} = \left( \text{erf}\left(\frac{\sqrt{2} r_{\text{safe}}}{w(Z_L)}\right) \right)^2.
\end{align}

Under the worst-case security assumption, we consider that any power falling outside this region is fully intercepted by the eavesdropper. Therefore, the pointing-related transmissivity to Eve is defined as:
\begin{align}
	\eta_{\text{po}}^{(E)} = 1 - \eta_{\text{safe}}, \quad \eta_E = \eta_{\text{po}}^{(E)}
\end{align}
Note that this partitioning is conditioned on the instantaneous pointing error. In security analysis, its statistical distribution can be integrated to evaluate expected leakage.

\section{Security Metric Analysis}
In this section, we analyze the effect of safe-zone, pointing error and turbulence on the achievable secret key rate (SKR) by examining the statistical behavior of the channel coefficients \( \eta_B \) and \( \eta_E \), and their impact on the information-theoretic quantities \( I_{AB} \) and \( \chi_{AE} \).

As previously modeled, atmospheric turbulence affects only Bob's channel and follows a Gamma-Gamma distribution. In contrast, pointing error is a shared random impairment that simultaneously impacts both Bob and Eve by determining the instantaneous angular deviation of the beam. Since $r_e^2 = Z_L^2 \left( \theta_{e_x}^2 + \theta_{e_y}^2 \right)$, and the sum \( \theta_{e_x}^2 + \theta_{e_y}^2 \) follows a chi-squared distribution with two degrees of freedom, the PDF of \( r_e^2 \) is thus given by:
\begin{align}
	f_{r_e^2}(r) = \frac{1}{2 Z_L^2 \sigma_{\theta}^2} \exp\left( -\frac{r}{2 Z_L^2 \sigma_{\theta}^2} \right), \quad r \geq 0
\end{align}

\subsection{Effective SKR and Mutual Information under Angular Thresholding}
To improve robustness against channel impairments and mitigate information leakage, we consider an angular-thresholding policy in which Bob accepts key generation only if the instantaneous radial misalignment satisfies \( r_e \leq r_{	ext{th}} \). This condition filters out high-deviation events that are likely to result in poor reception at Bob and stronger leakage to Eve. The threshold \( r_{\text{th}} \) thus directly controls the trade-off between mutual information and security.
Given the composite form of Bob's channel transmissivity:
\begin{align}
	\eta_B(r, h) = \eta_{\text{sys}} \cdot h \cdot A_0 \cdot \exp\left( -\frac{2r}{w^2(Z_L)} \right)
	\label{eq:eta_b_r_h}
\end{align}
and the eavesdropper's channel:
\begin{align}
	\eta_E(r) = 1 - A_{\text{safe}} \cdot \exp\left( -\frac{2r}{w^2(Z_L)} \right)
	\label{eq:eta_e_r}
\end{align}
we define the mutual information averaged over turbulence for a fixed radial offset \(r=r_e^2\) as:
\begin{align}
	&I(r) = \frac{1}{2} \int_0^\infty  f_{\eta_{\text{tur}}}(h) \nonumber \\
	&\log_2 \left( 1 + \frac{ \eta_{\text{sys}} \cdot h \cdot A_0 \cdot \exp\left( -\frac{2r}{w^2(Z_L)} \right) \cdot V_m }{ 1 - \eta_{\text{sys}} \cdot h \cdot A_0 \cdot \exp\left( -\frac{2r}{w^2(Z_L)} \right) + \xi } \right)  \, dh
	\label{eq:mutual_info_avg}
\end{align}
where $r<r_{\text{th}}^2$. This expression is only evaluated for misalignment values satisfying \( r=r_e^2 < r_{\text{th}}^2 \), in accordance with the angular filtering policy that discards all samples with excessive beam deviation. In effect, the threshold condition defines the support of integration over \( r_e^2 \) in all higher-level performance metrics.
The corresponding conditional key rate conditioned on $r=r_e^2$ is then given by:
\begin{align}
	K(r) = \beta I(r) - \chi_{AE}(\eta_E(r))
	\label{eq:skr_conditional_thresholded}
\end{align}
where $r<r_{\text{th}}^2$, and the Holevo bound term is:
\begin{align}
	\chi_{AE}(r) = \frac{1}{2} \log_2 \Bigg[ \frac{V_m + 1}{1 + \dfrac{ \left( A_{\text{safe}} \cdot \exp\left( -\frac{2r}{w^2(Z_L)} \right) \right) V_m }{2 - A_{\text{safe}} \cdot \exp\left( -\frac{2r}{w^2(Z_L)} \right) }} \Bigg]
\end{align}
By applying the threshold condition \( r_e \leq r_{\text{th}} \), we define the effective mutual information and SKR as:
\begin{align}
	\bar{I}(r_{\text{th}}) &= \frac{1}{2 Z_L^2 \sigma_{\theta}^2} \int_0^{r_{\text{th}}^2} I(r)  
	 \exp\left( -\frac{r}{2 Z_L^2 \sigma_{\theta}^2} \right) \, dr \\
	\bar{K}(r_{\text{th}}) &= \frac{1}{2 Z_L^2 \sigma_{\theta}^2} \int_0^{r_{\text{th}}^2} K(r) 
	  \exp\left( -\frac{r}{2 Z_L^2 \sigma_{\theta}^2} \right) \, dr
\end{align}
These metrics allow us to jointly assess the trade-off between system performance and security as a function of the angular acceptance threshold \( r_{\text{th}} \).

\begin{figure}
	\begin{center}
		\includegraphics[width=3.3 in, height= 1.9 in]{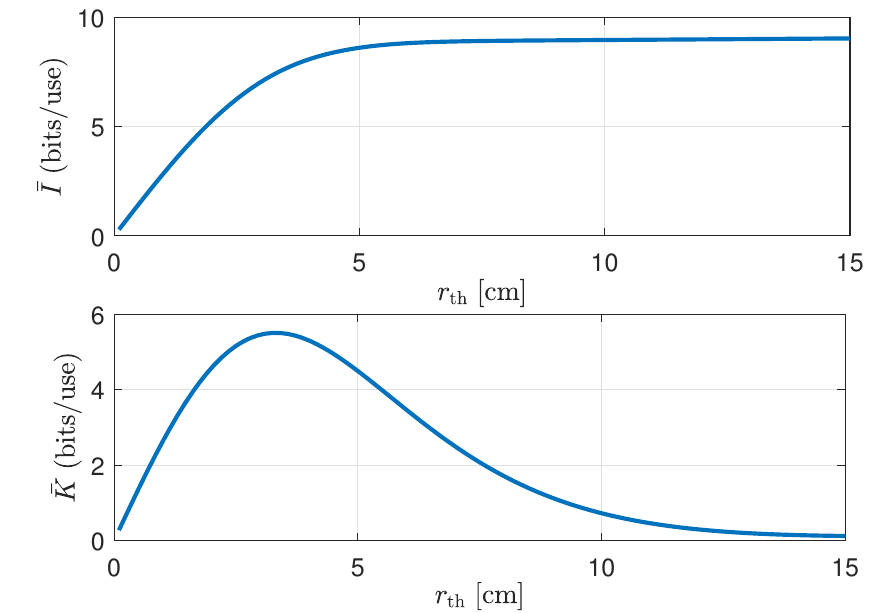}
		\caption{Average mutual information $\bar I$ and secret key rate $\bar K$ versus angular acceptance threshold $r_{\text{th}}$ for $Z_L=500$~m and $r_a=0.05$~m.}
		\label{sb1}
	\end{center}
\end{figure}

\begin{figure}
	\begin{center}
		\includegraphics[width=3.3 in, height= 1.8 in]{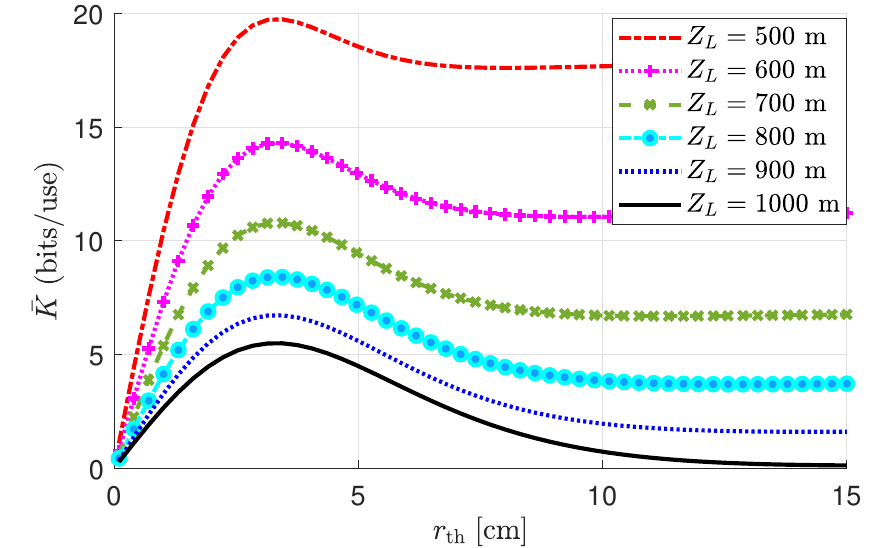}
		\caption{Average secret key rate $\bar K$ versus angular acceptance threshold $r_{\text{th}}$ for different link lengths $Z_L$ with $r_a=0.05$~m. Increasing $Z_L$ amplifies the impact of pointing errors, leading to lower key rates and reduced secure regions.}
		
		\label{sb2}
	\end{center}
\end{figure}

\section{Simulation Results and Discussion}
In this section, we investigate the impact of optimally tuning system parameters such as the beam waist at the receiver plane $w(Z_L)$ and the angular acceptance threshold $r_{\text{th}}$ on the achievable secret key rate (SKR) under different link conditions. Unless otherwise specified, the simulations are carried out using the default parameters $\eta_{\text{sys}}=0.8$, $r_a=0.05$~m, $r_{\text{safe}}=0.15$~m, $V_m=5$, $\xi=0.1$, $\beta=0.95$, $\sigma_{\theta}=50~\mu$rad, and $\lambda=1550$~nm, which are typical values adopted in the literature for continuous-variable free-space QKD systems. Variations in aperture radius $r_a$ or link length $Z_L$ are explicitly stated in the corresponding results.

\begin{figure*}
	\centering
	\subfloat[] {\includegraphics[width=3.4 in, height= 1.8 in]{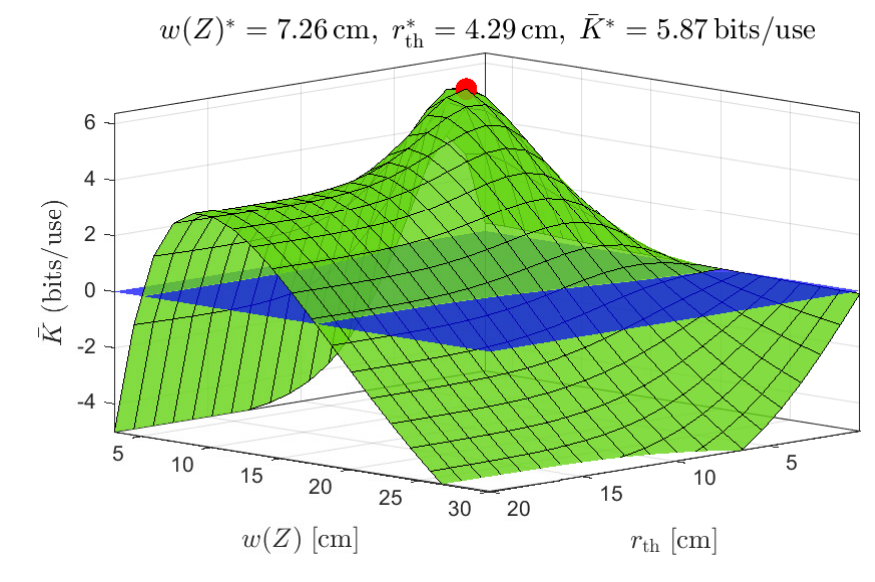}
		\label{cz1}
	}
	\hfill
	\subfloat[] {\includegraphics[width=3.4 in, height= 1.8 in]{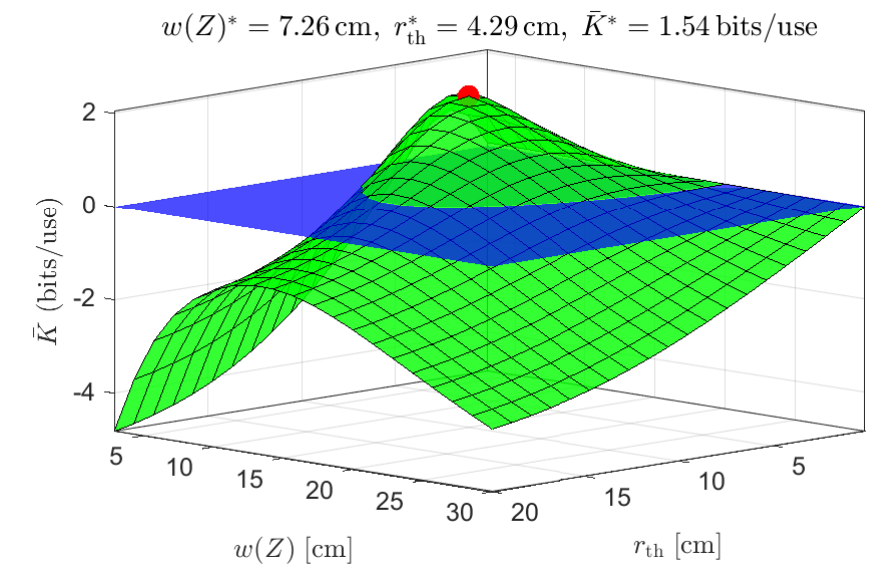}
		\label{cz2}
	}
	\hfill
	\subfloat[] {\includegraphics[width=3.4 in, height= 1.8 in]{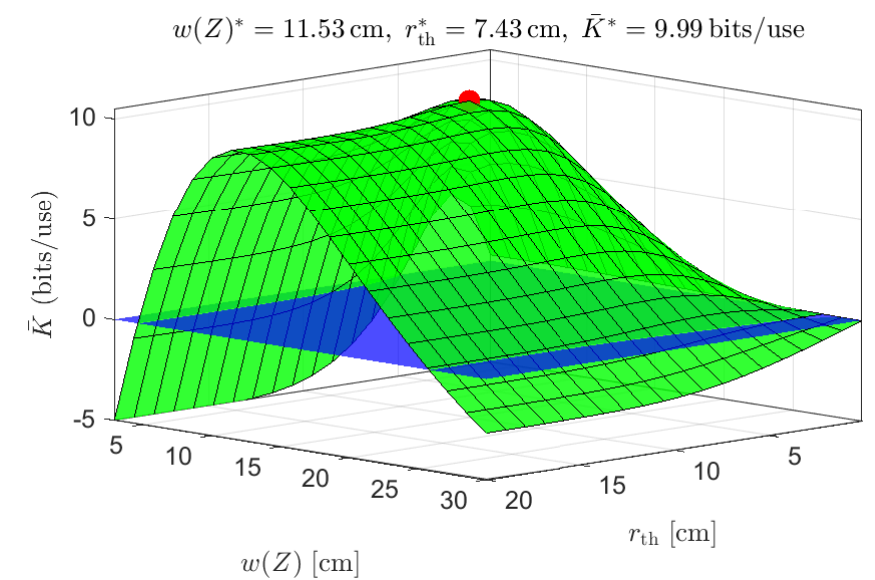}
		\label{cz3}
	}
	\hfill
	\subfloat[] {\includegraphics[width=3.4 in, height= 1.8 in]{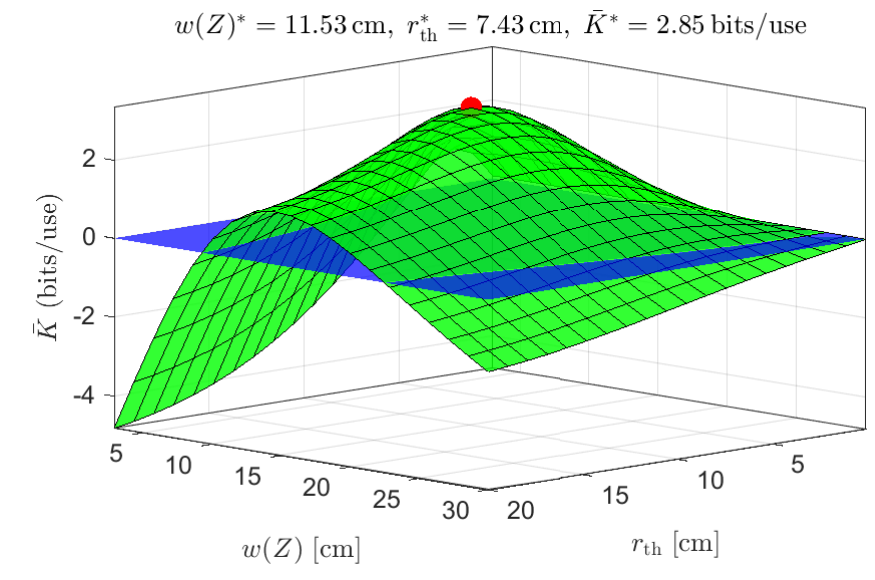}
		\label{cz4}
	}
	\caption{Three-dimensional plots of the average secret key rate $\bar K$ versus beam waist $w(Z_L)$ and angular threshold $r_{\text{th}}$ for two link lengths ($Z_L=500$ and $1000$~m) and two aperture radii ($r_a=0.05$ and $0.10$~m). }
	\label{cz}
\end{figure*}

Fig.~\ref{sb1} presents the average mutual information $\bar I$ and the average secret key rate $\bar K$ as functions of the angular acceptance threshold $r_{\text{th}}$ for the baseline case of $Z_L=500$~m and $r_a=0.05$~m. While $\bar I$ monotonically increases with $r_{\text{th}}$ because more symbols are retained, this does not necessarily translate into higher security. Larger thresholds also allow events with stronger pointing deviations, which substantially raise Eve's collected power and thereby her potential information gain. As a result, $\bar K$ follows a non-monotonic behavior: it increases initially, reaches an optimal peak, and then declines until vanishing at large $r_{\text{th}}$. This highlights the key trade-off in angular thresholding, simply maximizing $\bar I$ is insufficient, and the optimal $r_{\text{th}}$ must be chosen at the point where $\bar K$ is maximized to ensure both reliable reception and information-theoretic security.

Fig.~\ref{sb2} illustrates the average secret key rate $\bar K$ as a function of the angular acceptance threshold $r_{\text{th}}$ for different link distances $Z_L \in \{500,600,700,800,900,1000\}$~m with $r_a=0.05$~m. As the link length increases, the effect of angular jitter becomes more pronounced since small deviations translate into larger pointing displacements at the receiver plane. This results in stronger misalignment fading and higher leakage to Eve, which in turn reduces the achievable $\bar K$. Consequently, the peak value of $\bar K$ decreases with distance, and the overall secure region shrinks. These results emphasize that longer links require tighter angular control and carefully optimized thresholds to maintain security.

Fig.~\ref{cz} shows the three-dimensional behavior of the average secret key rate $\bar K$ as a function of both the beam waist at the receiver plane $w(Z_L)$ and the angular threshold $r_{\text{th}}$, for two link distances ($Z_L=500$ and $1000$~m) and two aperture sizes ($r_a=0.05$ and $0.10$~m). Each surface includes a reference plane at $K=0$, marking the boundary between secure and insecure operation, as well as the optimal point corresponding to the peak $\bar K$. The results reveal that enlarging the aperture radius has the strongest impact: it significantly increases the maximum achievable $\bar K$ and shifts the optimal operating point toward larger values of both $w(Z_L)$ and $r_{\text{th}}$, since a larger aperture tolerates wider beams and looser angular thresholds without excessive leakage. In contrast, extending the link distance mainly reduces the achievable peak value of $\bar K$ while leaving the location of the optimal $(w(Z_L), r_{\text{th}})$ nearly unchanged. Aperture size sets the balance between beam divergence and angular acceptance, while link distance mainly determines overall performance through channel losses and turbulence.

The simulation results emphasize the critical role of tuning system parameters to enhance security in continuous-variable free-space QKD. Adjusting the beam waist $w(Z_L)$ and the angular threshold $r_{\text{th}}$ allows balancing reliable reception against leakage to Eve. A larger aperture radius $r_a$ significantly improves the key rate but increases receiver size, which may be impractical for mobile platforms such as UAVs. Increasing the link distance $Z_L$ reduces the key rate due to stronger misalignment and turbulence, unless compensated by highly accurate pointing (well below $100~\mu$rad) at the cost of greater system complexity.


\enlargethispage{-0.22in}   
\makeatletter
\def\IEEEbibitemsep{0pt plus .2pt} 

\bibliographystyle{IEEEtran}
\bibliography{IEEEabrv,myref}

\end{document}